\documentclass[10pt,letterpaper,twocolumn]{article} 

\usepackage{ol2}
\usepackage[draft]{hyperref}
\usepackage{amsmath}

\begin{document}

\twocolumn[ 

\title{Observation of second-order hyper-Raman generation\\ in LiNbO$_3$ whispering-gallery mode disk resonators}


\author{Matt T. Simons,$^{*}$ and Irina Novikova}

\address{
College of William \& Mary, Williamsburg, VA. 23185, USA
\\
$^*$Corresponding author: mtsimons@email.wm.edu
}

\begin{abstract}
In this manuscript we report experimental demonstration of nonlinear frequency conversion at several optical frequencies in a whispering-gallery mode resonator (WGMR). Due to the enhancement of nonlinear interactions inside a WGMR, interaction of a $1064$~nm pump field with a LiNbO$_{3}$ disk produced a weak, but measurable non-phase matched $532$~nm second harmonic field at room temperature ($>100^\circ$~C below the phase-matching temperature), for pump powers of a few tens of mW. For higher pump powers, we observed the generation of four additional fields at $545$~nm, $559$~nm, $573$~nm, and $587$~nm. The relative spectral shift between two consecutive fields corresponds to a $455~\mathrm{cm}^{-1}$ vibrational mode in LiNbO$_{3}$ crystal. Our preliminary analysis indicates that these fields are the result of a multi-phonon hyper-Raman scattering, in which two photons of the pump field are converted into one photon of a higher-frequency field and one or several optical phonons.\\
\end{abstract}


 ] 


\noindent
Many applications, from a green laser pointer to an entangled photon source, take advantage of various frequency conversion processes in nonlinear  crystals. Typically, the weakness of nonlinear effects has to be compensated by use of a high-power pump laser field, high-quality optical cavities, or a combination of both. Whispering-gallery mode resonators (WGMRs), by supporting stable cavity modes of light traveling along the circumference of a disk or sphere through total internal reflection~\cite{OraevskyQuantElec2002,MatskoBook2009}, are a promising alternative to regular optical cavities formed by high-quality mirrors. WGMRs provide a unique combination of very high quality factors ($10^7 \sim 10^{12}$) and low mode volumes, which are unattainable with conventional optical cavities~\cite{BraginskyPL1989,MalekiOptExp2007}. As a result, WGMR applications span a wide range of fields, including communications, biophysics, and quantum optics~\cite{VahalaNature2003,ArmaniScience2007,MatskoBook2009}.

Enhancement of nonlinear conversion in WGMRs and their potential quantum optics applications have been a focus of several recent experiments. Dramatic improvement in the efficiency of nonlinear frequency conversion processes, including Raman scattering \cite{SpillaneNature2002,BumkiOptLett2003}, parametric oscillations \cite{SavchenkovOptLett2007,Furst2PRL2010}, and second \cite{IlchenkoPRL2004,Furst1PRL2010} and third \cite{SasagawaAppPhysExp2009} harmonic generation using WGMRs have been recently demonstrated. In this manuscript, we report observation of nonlinear conversion of a strong pump field ($\lambda_p=1064$~nm) into several optical fields of higher frequencies. One field, at $532$~nm, is the result of non-phase matched second harmonic generation (SHG). Each of the other four fields is consecutively offset by approximately $14$~nm from the SHG field or the closest higher-frequency generated field. Our analysis suggests that the best explanation for this phenomenon is second-order (multiphonon) hyper-Raman scattering --- a nonlinear process that, to the best of our knowledge, has not been previously demonstrated in a WGMR.

\begin{figure}[htb]
\centerline{\includegraphics[width=1.0\columnwidth]{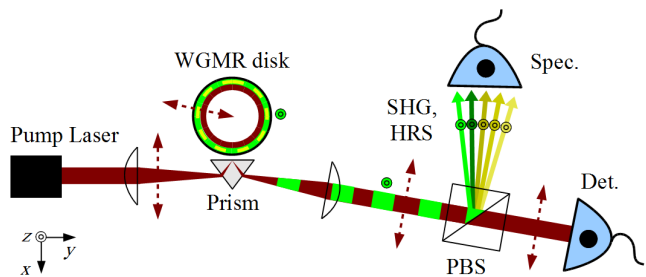}}
\caption{Schematic of the experimental apparatus (see text for abbreviations, arrows show polarizations).
}
\label{Fig:setup}
\end{figure}

A schematic of the experimental setup is shown in Fig.~\ref{Fig:setup}. Efficient coupling of laser radiation into and out of the WGMR is possible by focusing the horizontally polarized output of a $1064$~nm fiber laser on the internal surface of an equilateral diamond prism ($1.5$~mm sides), at the critical angle for total internal reflection between the prism and the disk. The coupling efficiency was adjusted by changing the separation between the prism and the disk~\cite{MatskoBook2009} using a precision micrometer.
In these experiments we used two $1$~mm-thick WGMR disks: one made of stoichiometric lithium niobate, and the other of magnesium oxide-doped lithium niobate (MgO:LiNbO$_3$, MgO $\sim 3.5\%$).  The diameters of both disks were approximately $7$~mm, corresponding to a free spectral range of $6$~GHz. Each disk was z-cut, with the extraordinary axis of the crystal perpendicular to the radius of the disk. The sides of the disks were hand-polished in our laboratory following the procedure developed in \cite{IlchenkoPRL2004}, to a rounded cross-section with approximately $0.2$~mm radius of curvature at the disk's equator.
The pump laser frequency had to be manually adjusted to a WGMR eigenmode to achieve maximum coupling. The lack of continuous frequency tuning capability significantly complicated the process of finding the optimal coupling and made the measurements of Q-factor at $1064$~nm wavelength nearly impossible. However, we estimate that the Q-factor exceeded $10^7$ from data obtained using a similar disk and a tunable $800$~nm laser.

The radiation inside the disk exited through the same coupling prism and was re-collimated by the second lens. All optical fields generated through nonlinear frequency conversion inside the disk were polarized in $z$ direction (as indicated in Fig.~\ref{Fig:setup}), orthogonal to the $x$-polarized pump laser. A polarizing beam splitter (PBS) placed after the output collimating lens split these fields from the transmitted pump laser and sent them to a broadband spectrometer (Spec.), with a resolution $0.4$~nm, not narrow enough to resolve individual WGMR modes. The transmission of the pump was used to control coupling efficiency.

\begin{figure}[htb]
\centerline{\includegraphics[width=0.8\columnwidth]{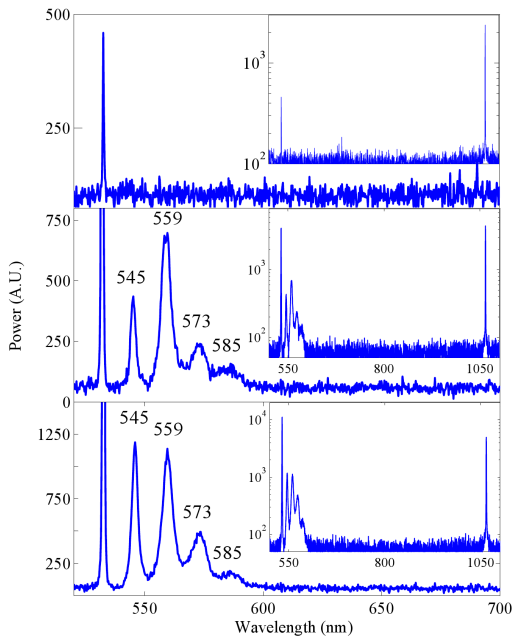}}
\caption{Spectra of emission from whispering-gallery mode resonators. (\emph{a}) SHG emission from stoichiometric LiNbO$_3$ disk with $11$~mW pump power; (\emph{b}) emission from same disk at $650$~mW pump power and coupling efficiency $70$\%; (\emph{c}) emission from MgO:LiNbO$_3$ disk at $825$~mW pump power and coupling efficiency $30$\%.}
\label{Fig:graphs}
\end{figure}

\begin{figure}[htb]
\centerline{\includegraphics[clip,width=0.6\columnwidth]{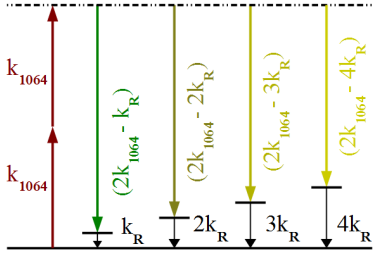}}
\caption{Photon schematics for hyper-Raman scattering and multi-phonon hyper-Raman overtones.}
\label{Fig:diagrams}
\end{figure}

To study nonlinear frequency conversion inside the disk, we gradually increased the pump laser power until we observed the generation of visible light. At room temperature, we observed second harmonic generation at $532$~nm inside our whispering-gallery mode disk  with a pump power of P$_{\omega} = 11$~mW, as shown in Fig.~\ref{Fig:graphs}(a). This result was somewhat surprising, as the experimentally determined phase matching temperature for our stoichiometric lithium niobate was T$_{PM} = 140~^{\circ}$C (MgO:LiNbO$_3$ T$_{PM} = 99~^{\circ}$C). The efficiency of the second harmonic conversion was significantly reduced due to the phase mismatch --- at a pump power of $P_{\omega} = 300$~mW the output second harmonic power was $P_{2\omega} \sim 3~\mu$W.
With increased pump power, we unexpectedly observed Raman-type generation of several additional fields, yellow-shifted with respect to the green $532$~nm SH field. In the stoichiometric lithium niobate disk [Fig.~\ref{Fig:graphs}(b)], these four unexpected fields appeared at roughly the same threshold power P$_{\omega} \simeq 430$~mW, at approximately equidistant frequencies $545$~nm, $559$~nm, $573$~nm, and $587$~nm. A similar spectrum was observed in MgO:LiNbO$_3$, at somewhat higher pump power due to less efficient coupling [Fig.~\ref{Fig:graphs}(c)].
The frequency position of the shifted modes, the scattering geometry inside the WGMR, and the relative intensity of the generated fields suggest that they are generated through a hyper-Raman process, accompanied by the excitation of one or multiple phonons, as shown in Fig.~\ref{Fig:diagrams}.

Selection rules for Raman and hyper-Raman transitions are determined by the relation between the polarizability and the symmetries of the phonon modes in a particular crystal lattice. The polarization response of a material to an external electromagnetic field can be generally expanded in the electric field to:
\begin{equation}
P_i = \alpha_{ij} E_j + \frac{1}{2} \beta_{ijk} E_j E_k + \frac{1}{6} \gamma_{ijkl} E_j E_k E_l + \ldots,
\end{equation}
where $\alpha$ is the polarizability, and $\beta$ and $\gamma$ are the first- and second- hyper-polarizability tensors, correspondingly~\cite{DenisovPhysRep1987,CyvinJChemPhys1965}. Notation for tensors $\alpha_{ij}$ and $\beta_{ijk}$ uses indices $i, j, k$ for the crystallographic axes $x$, $y$, $z$ with the first index $i$ corresponding to the polarization of the scattered radiation, and indices $j, k$ corresponding to the polarization of the incident photons. The scattering geometry for hyper-Raman and Raman scattering is written as d$_1$(ijk)d$_2$ and d$_1$(ij)d$_2$, where d$_1$ and d$_2$ are the directions of incident and scattered radiation, respectively. The scattering geometries in our WGMR were x(zyy)x and y(zxx)y for hyper-Raman, and x(zz)x and y(zz)y for Raman.

Lithium niobate has a trigonal crystal structure [space group C$_{3v}^6$(R3c)], and thus the optical phonon modes can be classified into 3 groups of modes: $4A_1 + 5A_2 + 9E$ [the number indicates the number of modes in each group, and each group can be further separated into transverse (TO) and longitudinal (LO)]~\cite{SchaufelePhysRev1966,RidahJPhysCondensMatt1997} . Each mode group couples to different polarization combinations of the incident and scattered fields, and thus each has different polarizability and hyperpolarizability tensors~\cite{SchaufelePhysRev1966,DenisovOptComm1978}.

\begin{figure}[htb]
\centerline{\includegraphics[clip,width=0.9\columnwidth]{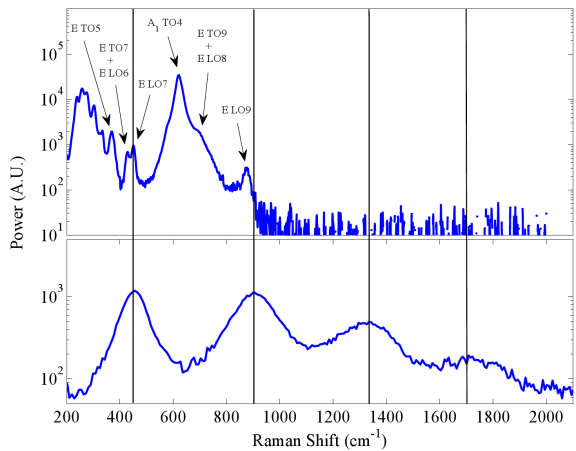}}
\caption{\emph{Top}: Raman scattering of $785$~nm from a lithium niobate crystal. \emph{Bottom}: Emission from lithium niobate WGMR with a $1064$~nm pump, referenced to $532$~nm.
Lines are at $455$, $909$, $1337$ and $1686$~cm$^{-1}$.}
\label{Fig:graphs2}
\end{figure}

The observed frequency difference between two consecutive generated fields of $\approx 14$~nm corresponds to a Raman shift of $\omega_{R} = 455$~cm$^{-1}$, associated with either the E(LO)7 or an A$_2$ phonon mode in LiNbO$_3$, as shown in Fig.~\ref{Fig:graphs2}~\cite{RidahJPhysCondensMatt1997,ClausPhysRev1972,HermetJCondMatt2007}. However, the generation of the yellow-shifted fields via direct Raman scattering of the second harmonic field of same polarization is prohibited by the selection rules, since the $\alpha_{zz}$ element of the E-mode polarizability tensor is zero. Similarly, we can rule out the A$_2$ mode as both its $\alpha_{zz}$ and $\beta_{zxx}$ elements are zero.
Rather, hyper-Raman scattering of the pump field is possible, since $\beta_{zxx}$ is nonzero for the E-modes.

Still, hyper-Raman scattering involving single phonons cannot explain the observation of the lower-frequency yellow fields. Such a process would involve the excitation of phonons with Raman shifts of $909$~cm$^{-1}$, $1337$~cm$^{-1}$, and $1686$~cm$^{-1}$, and these high-frequency phonons are not supported by the lithium niobate crystal lattice. This was verified by obtaining a traditional Raman scattering spectrum for our samples using a $785$~nm DeltaNu Raman Spectrometer. Comparison of this Raman spectrum with our WGMR emission is shown in Fig.~\ref{Fig:graphs2}.  We can also rule out cascaded Raman scattering as all the generated fields are polarized in the same $z$ direction, and the polarizability tensor element for that process is zero as described above. These extra fields are more consistent with multi-phonon overtones $2\omega_0 - 2\omega_R$, $2\omega_0 - 3\omega_R$, $2\omega_0 - 4\omega_R$ (Fig.~\ref{Fig:diagrams}), as they are beyond the first-order Raman spectrum and broaden with increasing wavenumber. Thus we think these results can be attributed to second-order hyper-Raman scattering.

In conclusion, we observed four Raman-shifted modes along with non-phase matched second harmonic generation inside our lithium niobate whispering-gallery mode disk resonators. The generated second harmonic field was neither powerful enough nor of the proper scattering geometry to produce observed Raman scattering at $455~$cm$^{-1}$. Thus, we attribute Stokes fields produced at $545$~nm, $559$~nm, $573$~nm, and $587$~nm to the hyper-Raman scattering and second-order hyper-Raman scattering of the $1064~$nm pump beam. This is interesting to study for both cavity hyper-Raman scattering itself, as well as its consequences for other nonlinear processes inside high-quality whispering-gallery mode resonators.

\bigskip

The authors thank A.~B. Matsko, A.~A. Savchenkov, and D.~V. Strekalov for valuable discussions and assistance, K.~L. Wustholz for the access to the commercial Raman spectrometer, S.~A. Aubin for the 1064 nm laser, and D.~J. Gribbin for polishing the disks. This research was supported by NSF grant PHY-0758010. M.~T.~S. acknowledges support from the Virginia Space Grant Consortium.

\end{document}